\documentclass[preprint,showpacs,preprintnumbers,amssymb,superscriptaddress,aps,prd,nofootinbib,11pt]{revtex4-1}
\usepackage{graphicx, epsfig, amssymb} 
\usepackage{amsmath, amsfonts}
\usepackage{bm} 
\usepackage{enumitem}
\usepackage[usenames]{color}
\definecolor{navyblue}{rgb}{0.0, 0.0, 0.5}
\usepackage[linktocpage,colorlinks=true,allcolors=navyblue]{hyperref}
\usepackage[caption=false]{subfig}
\usepackage[utf8]{inputenc}
\usepackage{tensor}
\usepackage{soul}

\def\beq{\begin{equation}}
\def\eeq{\end{equation}}

\def\bea{\begin{eqnarray}}
\def\eea{\end{eqnarray}}

\newcommand{\rmin}{\epsilon}
\newcommand{\rmax}{r_{\text{max}}}
\newcommand{\bmax}{b_{\text{max}}}
\newcommand{\EE}{\hat{E}}
\newcommand{\LL}{\hat{L}}
\newcommand{\Vgeo}{V_{\text{geo}}}
\newcommand{\Veff}{V_l^{\text{eff}}}
\newcommand{\Vwkb}{V_L^{\text{wkb}}}
\newcommand{\Aout}{A^+_l}
\newcommand{\Ain}{A^-_l}
\newcommand{\nn}{\nonumber}
\newcommand{\Thetageo}{\Theta_{\text{geo}}}
\newcommand{\Thetawkb}{\Theta_{\text{wkb}}}
\newcommand{\rinf}{r_{\infty}}
\newcommand{\bbow}{b_r}
\newcommand{\thetabow}{\theta_r}
\newcommand{\deltageo}{\delta_l^{\text{geo}}}
\newcommand{\fscat}{\hat{f}}

\begin{document}\title{\large Rainbow scattering in the gravitational field of a compact object}

\author{Sam R. Dolan}\email{s.dolan@sheffield.ac.uk}
\affiliation{Consortium for Fundamental Physics,
School of Mathematics and Statistics,
University of Sheffield, Hicks Building, Hounsfield Road, Sheffield S3 7RH, United Kingdom}

\author{Tom Stratton}\email{tstratton1@sheffield.ac.uk}
\affiliation{Consortium for Fundamental Physics,
School of Mathematics and Statistics,
University of Sheffield, Hicks Building, Hounsfield Road, Sheffield S3 7RH, United Kingdom}

\begin{abstract}
We study the elastic scattering of a planar wave in the curved spacetime of a compact object such as a neutron star, via a heuristic model: a scalar field impinging upon a spherically-symmetric uniform density star of radius $R$ and mass $M$. For $R < r_{c}$, there is a divergence in the deflection function at the light-ring radius $r_{c} = 3 \, GM/c^2$, which leads to spiral scattering (orbiting) and a backward glory; whereas for $R > r_{c}$ there instead arises a stationary point in the deflection function which creates a caustic and rainbow scattering. As in \emph{nuclear} rainbow scattering, there is an Airy-type oscillation on a Rutherford-like cross section, followed by a shadow zone. We show that, for $R \sim 3.5 \, GM/c^2$, the rainbow angle lies close to $180^\circ$, and thus there arises enhanced back-scattering and glory. We explore possible implications for gravitational wave astronomy, and dark matter models.
\end{abstract}

\date{\today}

\maketitle

\section{Introduction\label{sec:intro}}

The era of gravitational wave astronomy began in 2015 with the first direct detection of gravitational waves (GWs) \cite{Abbott:2016blz, TheLIGOScientific:2016qqj}. This era promises rich new data on the strong-field dynamics of compact objects such as black holes and neutron stars \cite{Shapiro:1983du}. The ``chirp'' signal GW150914, observed at Advanced LIGO, seems to be in accord with the predictions of Einstein's general relativity for a binary black hole merger, as modelled through post-Newtonian theory and numerical relativity \cite{TheLIGOScientific:2016pea, Buonanno:2006ui}. The `ringdown' in GW150914 suggests that the merger product is a Kerr black hole \cite{TheLIGOScientific:2016src, TheLIGOScientific:2016htt, Yunes:2016jcc, Chirenti:2016hzd}. New data promises to further constrain the window for alternatives \cite{Cardoso:2016rao,Konoplya:2016pmh}.  

Gravitational-wave astronomy is complementary to electromagnetic astronomy, in part, because GWs penetrate the shrouds of dust and gas that typically obscure the most energetic parts of the universe. Yet, GWs may still be scattered indirectly, by the influence of matter/energy on the curvature of spacetime. 
In principle, the scattering of GWs provides information on the strong-field geometry of compact objects (i.e.~black holes, neutron stars and white dwarfs), or hypothetical exotic alternatives (e.g.~boson stars \cite{Liebling:2012fv,Macedo:2013jja}, `hairy' black holes \cite{Herdeiro:2014goa, Cunha:2015yba}, or wormholes \cite{Damour:2007ap,Konoplya:2016hmd}).
  
The time-independent scattering of planar GWs (and other fundamental fields) by black holes has been the subject of numerous works since 1968 \cite{Matzner:1968,Sanchez:1977vz,Handler:1980un,Zhang:1984vt,Matzner:1985rjn,Futterman:1988ni,Anninos:1992ih,Andersson:2000tf,Glampedakis:2001cx,Dolan:2007ut,Dolan:2008kf,Crispino:2009ki,Crispino:2009xt,Batic:2012rm,Kanai:2013rga,Crispino:2014eea,Crispino:2015gua,Sorge:2015yoa,Gussmann:2016mkp,Cotaescu:2016aty,Macedo:2016yyo}. By comparison, the time-independent scattering of GWs by compact bodies such as neutron stars has received little attention (though for related work see e.g.~\cite{Tominaga:1999iy,Tominaga:2000cs,Bernuzzi:2008rq}). 

It is well-established -- theoretically, at least -- that a typical black hole scattering cross section $d\sigma / d\Omega$ exhibits `spiral scattering' (orbiting) oscillations \cite{Anninos:1992ih} and a backward `glory' \cite{Matzner:1985rjn} (see e.g.~Fig.~10 in Ref.~\cite{Dolan:2008kf}). These effects may be understood in terms of the properties of the deflection function $\Thetageo(b)$ for the null geodesics (`rays') of the spacetime. For a Schwarzschild black hole of mass $M$, the deflection function \emph{diverges} at the critical impact parameter $b_c = \sqrt{27} GM/c^2$, where $G$ is the gravitational constant and $c$ is the speed of light. The ray with $b=b_c$ asymptotes towards the light-ring at radius $r_c = 3 G M /c^2$. Rays with $b > b_c$ are scattered, whereas rays with $b < b_c$ pass into the event horizon of the black hole. Due to the divergence, there exists (in principle) a ray scattered through \emph{any} arbitrary angle. Heuristically, the interference between rays passing through angles $\theta$, $2\pi - \theta$, $2\pi + \theta$, etc., gives rise to orbiting, and the interference between rays scattered near $\pi$, $2\pi$, $3\pi$, etc., gives rise to glories \cite{Ford:1959, Ford:1959b}.

\begin{figure}
\subfloat[The rainbow parameters $\bbow$ and $\thetabow$]{\label{subfig:sketch}%
  \includegraphics[width=0.45\textwidth]{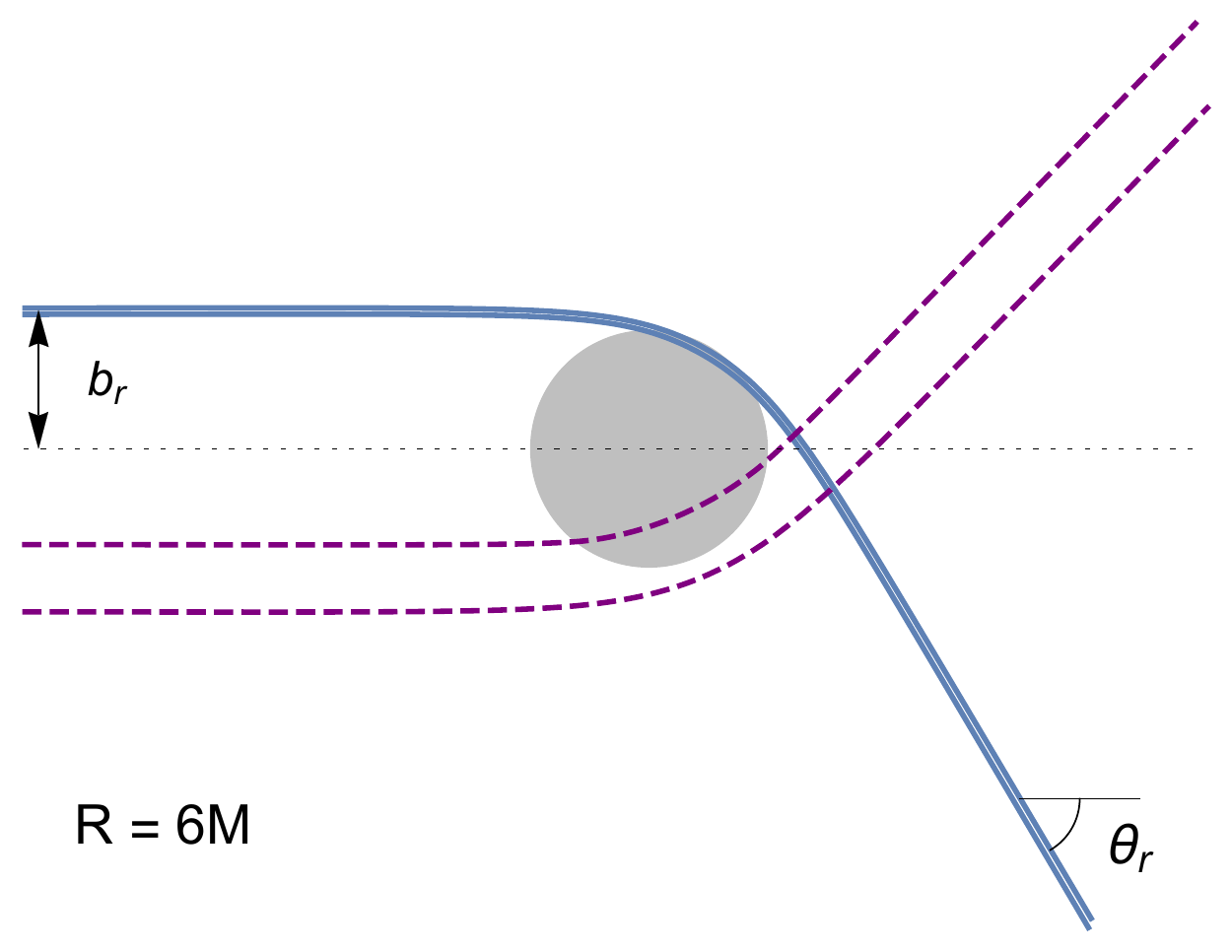}%
}
\subfloat[The caustic and rainbow wedge]{\label{subfig:caustic}%
  \includegraphics[width=0.45\textwidth]{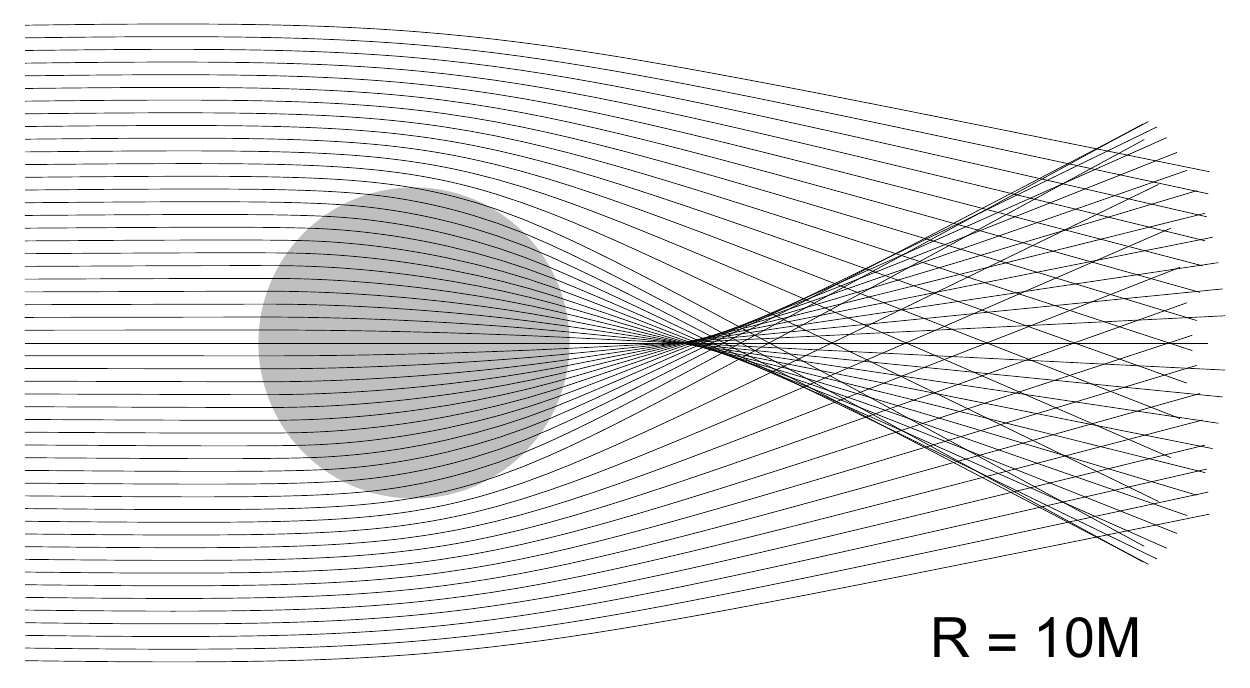}%
} \\
\subfloat[The deflection function]{\label{subfig:deflection}%
  \includegraphics[width=0.7\textwidth]{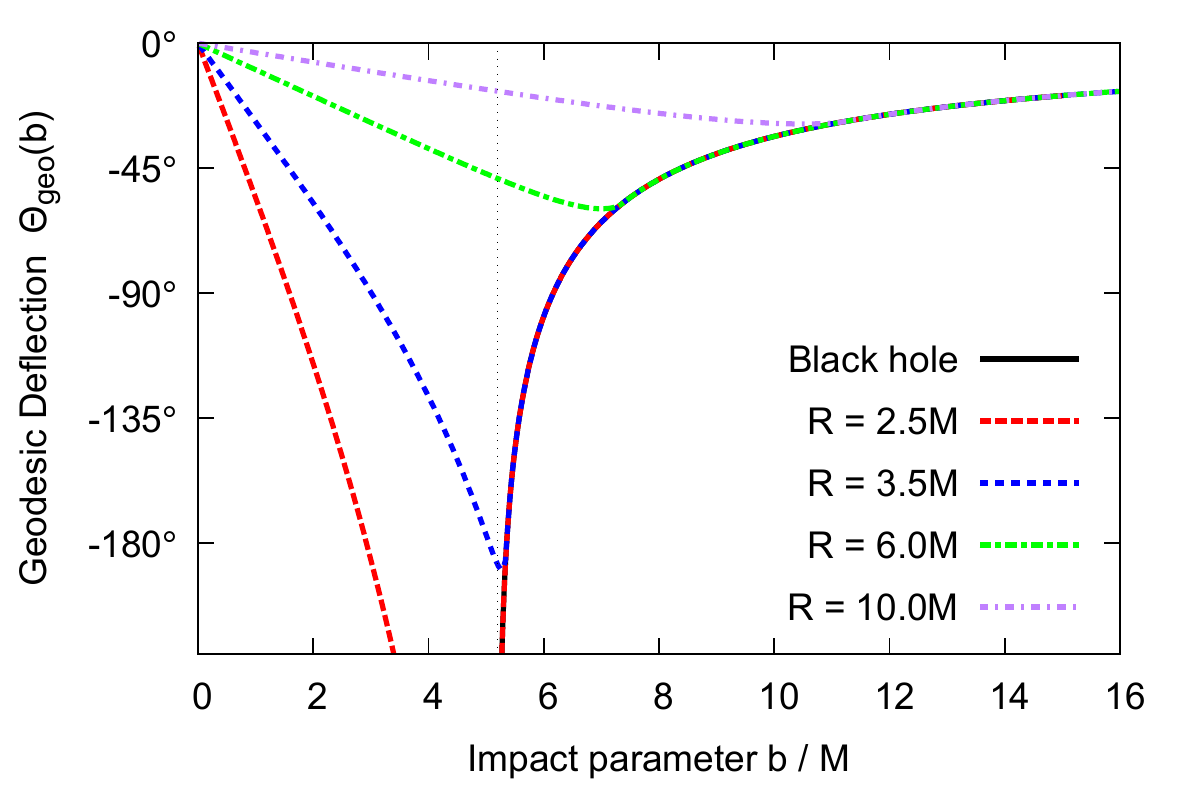}%
}
\caption{
(a) Null geodesics (rays) scattered by the spacetime curvature of a compact body of radius $R = 6M$ (N.B.~$G=c=1$). The solid (blue) lines show neighbouring rays with impact parameter near $\bbow$ which are scattered through the rainbow angle $\thetabow$. The dashed (purple) lines show another pair of geodesics which are both scattered through an identical angle $\theta$ (with $\theta < \thetabow$), generating an interference effect. 
(b) Rays passing through a compact body of radius $R = 10M$, generating a caustic behind the body, and a scattered `wedge' within $\theta_r$. 
(c) The geodesic deflection function $\Thetageo(b)$. By convention, the deflection is negative for attractive interactions. For black holes, or for highly compact bodies with $R < r_c = 3M$, there is a divergence at $b_c = \sqrt{27} M$, associated with a light-ring. Conversely, for $R > r_c$ there is a stationary point in the deflection function, $\Thetageo^\prime(\bbow) = 0$, leading to rainbow scattering.  For the cases $R = 10M$, $6M$ and $3.5M$, the rainbow angles are $\thetabow = 29.1^\circ$, $59.6^\circ$, $189.4^\circ$, respectively. 
}
\label{fig:deflection}
\end{figure}

By contrast, in compact bodies of radius $R > r_c$ a light-ring is not extant, and neither is an absorbing horizon. Instead of a divergence, the deflection function $\Thetageo(b)$ will (generically) possess one or more stationary points (see Fig.~\ref{fig:deflection}). A stationary point arises in any deflection function which is sufficiently smooth, and which has the reasonable asymptotic properties $\Thetageo(b) \sim -4 GM / (c^2 b)$ (the Einstein deflection angle)\footnote{By convention, $\Theta(b)$ is negative for attractive scattering} in the weak field ($b \gg GM/c^2$) and $\Thetageo(0) = 0$ for head-on collisions. As illustrated in Fig.~\ref{subfig:caustic}, a stationary point typically generates a ray pattern with a caustic and a rainbow wedge. 

In scattering processes, the classical scattering cross section, 
\beq
\left. \frac{d \sigma}{d \Omega} \right|_{\text{cl}} = \frac{b}{\sin \theta \, \left| \frac{d \Theta}{d b} \right|} ,
\label{eq:classical}
\eeq
is singular at the poles ($\theta = 0$, $\pi$) and it is also singular at stationary points of the deflection function, where $\Theta^\prime(\bbow) = 0$. In semiclassical theory, starting with Airy's work of 1838 \cite{Airy:1838,Ford:1959}, the singularities are transmuted into familiar interference effects: the former become glories \cite{Laven:2005}, and the latter lead to rainbow scattering oscillations.

In rainbow scattering \cite{Ford:1959}, the condition $\Theta^\prime(\bbow) = 0$ defines a rainbow impact parameter $\bbow$, a rainbow angle $\thetabow \equiv |\Theta(\bbow)|$, and a second derivative $\Theta_r'' \equiv \tfrac{d^2\Theta}{db^2}(\bbow)$. Airy's work \cite{Airy:1838} brought insight into two key features of the rainbow which are familiar from everyday meteorological experience: (1) the colours of the rainbow are separated in angle according to wavelength, with the `primary' peak appearing at $ \thetabow - 0.237 [\lambda^2 \Theta_r'']^{1/3}$ (see Eq.~(\ref{eq:airy})), where $\lambda$ is the wavelength; and (2) on the bright side of the rainbow the cross section has supernumerary peaks beyond the primary, whereas on the dark side of the rainbow the intensity falls off rapidly in the classically-forbidden shadow region.

Herein, we shall describe an effect somewhat akin to \emph{nuclear rainbow scattering} in the collisions of ions \cite{Khoa:2006id}. Ion-scattering experiments \cite{Goldberg:1974zza, Khoa:2006id} measure cross sections in which rainbow oscillations appear in conjunction with a Rutherford-like background arising from the long-ranged Coulomb interaction. In fact, in the nuclear case the quantum-mechanical deflection function possesses \emph{two} stationary points, linked to the (repulsive) Coulomb interaction and the (attractive) nuclear interaction, respectively: see Fig.~9 in \cite{Khoa:2006id}. The former leads to small-angle rainbow scattering and the latter to the wide-angle rainbow features that were first observed in the 1970s \cite{Goldberg:1972zzb,Goldberg:1974zza} (see e.g.~Fig.~11 in \cite{Khoa:2006id}). From the latter, one may seek to infer the properties of the nuclear potential. 

Our purpose here is to explore the general features of rainbow scattering from the spacetime geometry of compact objects. We are content to study a heuristic model: a scalar field propagating on a spherically-symmetric curved spacetime made by a uniform-density star (or other matter distribution). The key feature of our model is that the wave is scattered by geometry only: the wave may pass into and out of the compact body without significant attenuation or absorption. This feature is shared by a gravitational wave, and -- hypothetically -- by certain ultra-light dark matter candidates \cite{Jaeckel:2010ni,Arias:2012az}. Our simple model is reviewed with a critical eye in Sec.~\ref{sec:discussion}. 

Henceforth, we adopt the standard convention $G = c = 1$, so that (e.g.) $R/M = R c^2 / GM$ represents a dimensionless `compactness parameter'. Some characteristic values for $R/M$ include $R/M \sim 6$ for neutron stars, $\sim 1.4 \times 10^3$ for a massive white dwarf (e.g.~Sirius B), $\sim 9.4\times10^3$ for a typical white dwarf, $4.7 \times 10^5$ for the Sun, and $1.4 \times 10^9$ for Earth. 

In Sec.~\ref{sec:model} we describe the model and methods, addressing the spacetime and its geodesics; the partial-wave approach; and aspects of semiclassical theory that lead to Airy's rainbow formula. In Sec.~\ref{sec:results} we present a selection of numerical results for the scattering cross section $d\sigma / d\Omega$ in our  model, and we describe the key features. We conclude with a discussion of physical implications and open questions in Sec.~\ref{sec:discussion}.

%\cite{Macedo:2013jja,Shapiro:1983du,Ashton:2016xff,Goldberg:1972zzb,Abedi:2016hgu,Sanchez:1977vz,Crispino:2009xt,Matzner:1985rjn,Handler:1980un,Cardoso:2016rao,Cardoso:2014sna,Khoa:2006id,Anninos:1992ih,Batic:2012rm,Goldberg:1974zza,Dolan:2008kf,Andersson:2000tf,Futterman:1988ni,Crispino:2015gua,Dolan:2007ut,Crispino:2009ki,Marinho:2016ixt,Glampedakis:2001cx,Ford:1959,Dolan:2016bxj,Hod:2017xkz,Zhang:1984vt}

\section{Model and methods\label{sec:model}}
\subsection{Spacetime geometry and geodesics\label{subsec:geodesics}}

We take as our gravitating source a spherically-symmetric incompressible perfect fluid ball of uniform density \cite{Shapiro:1983du} in a coordinate system $\{t,r,\theta,\phi\}$. The line element is
\beq
ds^2 = g_{a b} dx^{a} dx^{b} = - f dt^2 + h^{-1} dr^2 + r^2 d\Omega^2 , 
\eeq
where $d\Omega^2 = d\theta^2 + \sin^2\theta d\phi^2$. 
The radial function $h(r)$ is continuous but not differentiable across the surface of the star at $r = R$ and the radial function $f(r)$ is once, but not twice, differentiable. In the exterior $r>R$, we have $f(r) = h(r) = 1 - 2M/r$, by Birkhoff's theorem \cite{VojeJohansen:2005nd}. In the interior $r<R$, we have \cite{Shapiro:1983du}
\bea
f(r) &=& \frac{1}{4R^3} \left( \sqrt{R^3 - 2Mr^2} - 3 R \sqrt{R - 2M} \right)^2 , \nn \\
h(r) &=& 1 - \frac{2Mr^2}{R^3} .
\eea

This is known as Schwarzschild's interior solution for an incompressible fluid \cite{Schwarzschild:1916}. Schwarzschild showed that, for the pressure to be finite at the origin, the bound $R/M > 9/4$ must be satisfied. Buchdahl \cite{Buchdahl:1959zz} showed that the same bound applies to {\it any} perfect fluid sphere with a monotonically decreasing density $\rho(r)$ and a barotropic equation of state. `Buchdahl's bound', as it is commonly known, was strengthened to $R/M > 8/3$ by including the dominant energy condition \cite{Barraco:2002ds}, and to $R/M \gtrsim 2.74997$ by demanding a sub-luminal speed of sound \cite{Fujisawa:2015nda}. Recent observations of recycled pulsars suggest that neutron stars can reach a mass of $M \sim 2 M_{\odot}$ and radius $R \sim 10\text{km}$ \cite{Ozel:2016oaf}, giving a ratio of $R/M \sim 3.4$.

Geodesics are extremal paths $x^a(\nu)$ of the action $S = \int \mathcal{L} \, d\nu$ with Lagrangian $\mathcal{L} = \tfrac{1}{2} g_{ab} \dot{x}^a \dot{x}^b$, where $\dot{x}^a = dx^a / d\nu$ with $\nu$ an affine parameter. Without loss of generality, we may restrict attention to motion in the equatorial plane. The Euler-Lagrange equations give two constants of motion, $\EE \equiv f \dot{t}$ and $\LL \equiv r^2 \dot{\phi}$.  The impact parameter $b$ is defined by their ratio $b \equiv \LL / \EE$. The null condition $\mathcal{L} = 0$ yields an `energy equation',
\beq
\dot{r}^2 = h f^{-1} \left(\EE^2 - \Vgeo(r) \right), \quad \quad \quad \Vgeo(r) \equiv f \LL^2 / r^2 .
\eeq

\begin{figure}
\includegraphics[width=10cm]{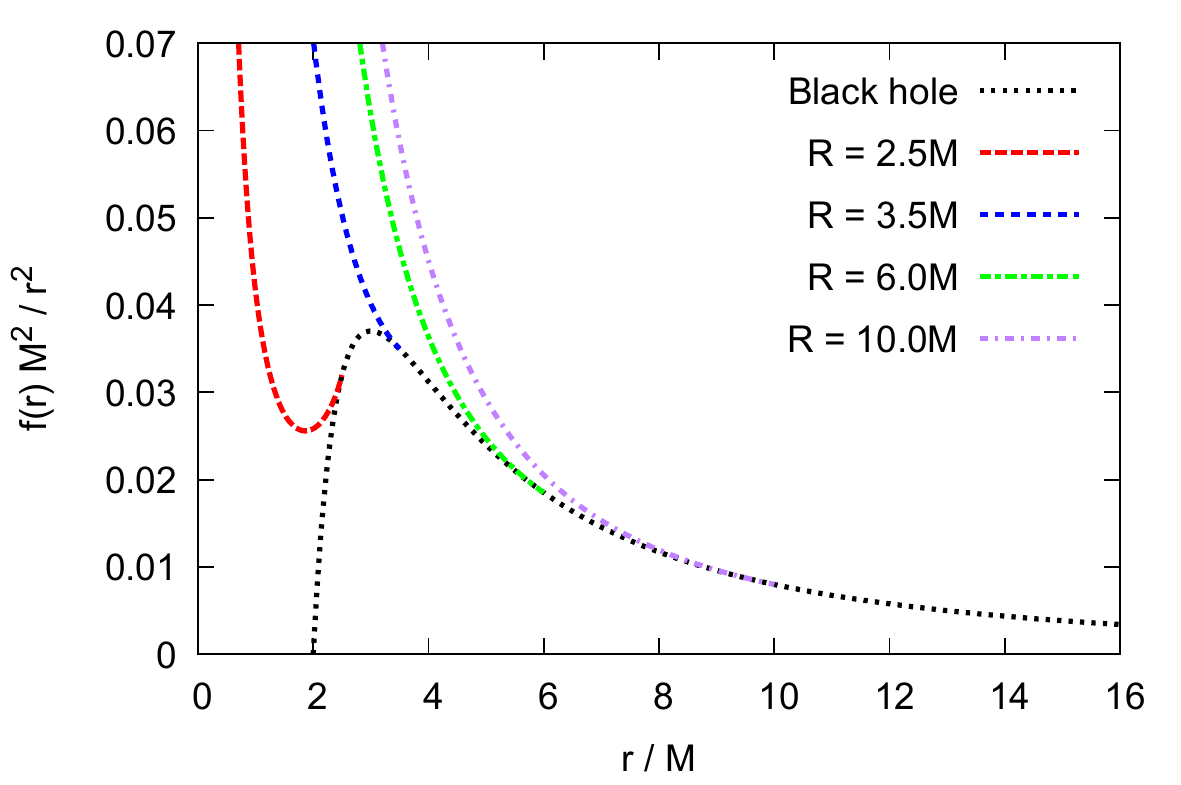}
\caption{
Null geodesic potential $\Vgeo / \LL^2 = f(r) / r^2$, for a black hole and for compact bodies of radius $R = 2.5M$, $3.5M$, $6M$ and $10M$. %The closest approach of a ray is found from $\Vgeo(r) = \EE^2$.
}
\label{fig:potential}
\end{figure}
 Figure \ref{fig:potential} illustrates that the potential $\Vgeo(r)$ has no stationary points for $R > r_c = 3M$; one stationary point for $R = r_c$; and two for $R < r_c$. In the latter case, the stationary points correspond to (outer) unstable and (inner) stable circular null orbits. We note in passing that stable null orbits give rise to intriguing phenomena \cite{Dolan:2016bxj} including instabilities \cite{Cardoso:2014sna}.

As described in Sec.~\ref{sec:intro} the geodesic deflection angle $\Thetageo(b)$ provides rudimentary insight into the scattering process. The deflection $\Thetageo$ is found by integrating $d\phi/dr = \sqrt{\dot{\phi}^2 / \dot{r}^2}$ along a null ray, to obtain
\beq
\Thetageo = \pi - 2 \int_{r_0}^\infty \frac{\LL}{r^2 \sqrt{ \EE^2 - \Vgeo(r)} } \sqrt{\frac{f}{h}} dr,  \label{eq:thetageo}
\eeq
where $r_0$ is the turning point of radial motion, satisfying $\Vgeo(r_0) = \EE^2$. By convention, $\Thetageo$ is negative for attractive interactions. Furthermore, as the deflection angle depends only on the ratio $b = \LL/\EE$ we may write $\Thetageo(b)$ without ambiguity. 

In Schwarzschild spacetime, the deflection angle can be written in terms of elliptic integrals. In the compact-body spacetime we are content to compute $\Thetageo(b)$ numerically. This may be done by evaluating the integral (\ref{eq:thetageo}) by quadrature; or, one may solve the Euler-Lagrange equations with numerical methods.  In the latter case, one may start with initial conditions $\phi(0) = 0$, $r(0) = \rinf$, 
\beq
\dot{r}(0) = -\left. \sqrt{h f^{-1} \left(\EE^2 - \Vgeo(r) \right)} \right|_{r=\rinf},
\eeq 
and integrate the geodesic equations up to $\nu_{\infty} (>0)$ defined by $r(\nu_\infty) = \rinf$, choosing some suitably large value for $r_{\infty}$. % We chose $\rinf = 10^4 M$. 

The geodesic deflection function $\Thetageo(b)$ is shown in Fig.~\ref{fig:deflection}. It has one stationary point if $R > 3M$, and no stationary points otherwise. In the latter case it exhibits a divergence at $b_c= \sqrt{27}M$.

\subsection{Waves and scattering}

We consider a massless scalar field $\Phi$ governed by the Klein-Gordon equation on the curved spacetime, viz.
\beq
\nabla_\mu \nabla^\mu \Phi = \frac{1}{\sqrt{-g}} \partial_\mu \left( \sqrt{-g} g^{\mu \nu} \partial_\nu \Phi \right) = 0 ,
\eeq
where $g^{\mu \nu}$ is the inverse metric and $g$ is the metric determinant. Employing the separation of variables method with the standard ansatz,
\beq
\Phi = \frac{1}{r} \sum_{lm} a_{lm} e^{-i \omega t} u_l(r) Y_{lm}(\theta, \phi),
\eeq
leads to the radial equation
\beq
\left[ \frac{d^2}{dr_\ast^2} + \omega^2 - \Veff(r) \right] u_{l}(r) = 0, \label{eq:radial}
\eeq
with an effective potential
\beq
\Veff(r) = f \left( \frac{l (l+1)}{r^2} + \frac{h}{2r} \left(\frac{f'}{f} + \frac{h'}{h} \right) \right) . \label{eq:Veff}
\eeq
Here we have introduced a tortoise coordinate defined by $dr / dr_\ast = \sqrt{f h}$. 
The modes $u_l(r) / r$ are required to be regular at the origin of the compact body. %, in the case of black holes, purely ingoing at the event horizon. 

We use standard time-independent scattering theory, as developed in Refs.~\cite{Newton:1982qc,Landau:Lifshitz}, and as described in the black hole context in Ref.~\cite{Frolov:1998wf}. The scattering cross section $d\sigma / d\Omega$ -- our primary object of interest -- is the square modulus of the scattering amplitude $\fscat(\theta)$, where
\beq
\fscat(\theta) = \frac{1}{2i\omega} \sum_{l=0}^\infty (2 l + 1) (S_l - 1) P_l(\cos \theta) . \label{eq:fscat}
\eeq 
Here $P_l(\cdot)$ are Legendre polynomials. The scattering coefficient $S_l $ may be expressed in terms of a phase shift, $S_l \equiv \exp(2 i \delta_l)$. It is determined from
\beq
S_l = (-1)^{l+1} \Aout / \Ain,  \label{eq:S}
\eeq
where $\Aout$ and $\Ain$ are the outgoing and ingoing coefficients of the mode in the far-field ($r \rightarrow \infty$),
\beq
u_l(r) \sim  \Aout e^{i \omega r_\ast} + \Ain e^{-i \omega r_\ast} .
\eeq
%The scattering coefficients $S_l$ may be written in terms of phase shifts $\delta_l$, via $S_l = \exp(2 i \delta_l)$.

\subsection{Semiclassical methods}
The standard semiclassical prescription \cite{Ford:1959} involves introducing several approximations into Eq.~(\ref{eq:fscat}). First, a WKB approximation for the phase shift $\delta_l$. Second, large-$l$ asymptotics for Legendre polynomials,
\beq
P_l(\cos \theta) \approx \begin{cases} \left(\tfrac{1}{2} L \pi \sin \theta\right)^{-1/2} \sin\left(L \theta + \pi/4\right), \quad & L \sin \theta \gtrsim 1, \\ 
(\cos \theta)^{L-1/2} J_0( L \theta ), & L \sin \theta \lesssim 1,  \end{cases}
\eeq
where $L \equiv l + \tfrac{1}{2}$. Third, replacing the sum over $l$ with an integral; and fourth (optionally)  evaluating the integral with the method of stationary phase, or other suitable method.

Let us examine the first step in more detail. In place of the phase shift $\delta_l$ we may use the WKB approximation,
\beq
\delta_l^{\text{wkb}} =  \tfrac{1}{2} L \pi - \omega r_0^\ast + \int_{r_0^\ast}^\infty \left\{ \sqrt{\omega^2 - \Vwkb(r)} - \omega \right\} dr_\ast ,
\eeq
with $r^\ast_0 \equiv r_\ast(r_0)$ the turning point defined by $\omega^2 =  \Vwkb(r_0)$. Here we have made the usual `Langer replacement' \cite{Langer:1937qr} $l (l+1) \rightarrow L^2$ to obtain $\Vwkb$ from the effective potential $\Veff$ defined in Eq.~(\ref{eq:Veff}). The wave-scattering deflection function $\Theta(L)$ is defined by 
\beq
\Theta(L) \equiv \frac{d \left(2 \delta_l\right)}{dL} .  \label{eq:deflectiondef}
\eeq 
In this expression, $L \equiv l+1/2$ is allowed to take real values. Inserting the WKB expression for the phase shift into (\ref{eq:deflectiondef}), and using $dr_\ast = dr / \sqrt{f h}$, yields
\beq
\Thetawkb(L) = \pi - 2 \int_{r_0}^\infty \frac{L}{r^2 \sqrt{\omega^2 - \Vwkb(r)}} \sqrt{\frac{f}{h}} \,  dr.  \label{eq:thetawkb}
\eeq
Now we see that Eq.~(\ref{eq:thetawkb}), for $\Thetawkb$, takes the same form as Eq.~(\ref{eq:thetageo}), for $\Thetageo$. Thus, in the semiclassical picture we may associate a partial wave with a null geodesic of impact parameter $b = (l+1/2) / \omega$. With this mapping, it follows that $\Vwkb / \omega^2 = \Vgeo / \EE^2 \left(1 + O(b^{-2}) \right)$ and thus we may expect $\Thetawkb$ to be approximately equal to $\Thetageo$ when $b \gg M$. % (There is one important caveat: the WKB approximation is not strictly valid if $\Veff(r_0) = \omega^2$ has more than one root, unless the roots are sufficiently well-separated; thus it is of questionable validity near the peak of a potential barrier).

Interference effects arise when more than one ray emerges at the same scattering angle. If two such rays with impact parameters $b_j = L_j/\omega$ are sufficiently separated ($|b_1 - b_2| \omega \gg 1$) then their semi-classical contributions to the integral may be treated separately using the method of stationary phase, leading to
\begin{align}
\frac{d\sigma}{d\Omega} &\approx %\frac{1}{ \omega^2 \sin \theta} \bigg( \sum_{j=1}^2  \frac{L_j}{| \Theta'_j |} + 2  \sqrt{ \frac{L_1 L_2}{\Theta'_1 \Theta'_2} } \cos \big( \phi_+^1 - \phi_+^2 \big)   \bigg) \nonumber \\ & \approx 
\sigma_{cl}^1 + \sigma_{cl}^{2} + \sqrt{\sigma_{cl}^{1}\sigma_{cl}^{2}}\cos \big[ 2(\delta^1 - \delta^2) + (L_1 - L_2)\theta \big],
\end{align}
where $\sigma_{cl}^j$ are the classical cross sections of Eq.~(\ref{eq:classical}), and $\delta^j$ are constant phase terms. The interference effect leads to regular oscillations in the cross section with the angle $\theta$. 

If the rays are not well-separated, for example, if the rays are on either side of a stationary point in $\Theta(b)$ as shown in Fig.~\ref{fig:deflection}(a), more care is needed. In Ref.~\cite{Ford:1959} it is shown that applying the standard semiclassical prescription to the case where $\Thetawkb$ possesses a stationary point leads to Airy's formula \cite{Airy:1838},
\beq
\frac{d\sigma}{d\Omega} \approx \frac{2 \pi \bbow}{\omega q^{2/3} \sin \theta} \text{Ai}^2\left[ (\theta - \thetabow) q^{-1/3} \right], \quad \quad q \equiv \frac{\Theta''_r}{2 \omega^2} ,  \label{eq:airy}
\eeq
where $\text{Ai}(\cdot)$ is the Airy function of the first kind. 
If $M \omega$ is sufficiently large then we may insert geodesic values for $\bbow$, $\thetabow$ and $\Theta_r''$ into the above approximation. We shall verify this in Sec.~\ref{subsec:airy}.

\subsection{Computational methods}

 \subsubsection{Scattering coefficients\label{subsec:numerical}}
 We used Mathematica to calculate scattering coefficients $S_l$ with the following approach: 
 \begin{enumerate}[itemsep=-1mm]
  \item start at $r=\rmin$ with initial values $u_l(\rmin)$ and $u_l'(\rmin)$, determined from the regular Frobenius series up to $O(r^{l + n_0})$; 
  \item find numerical solutions for $u_l(r)$ and $u_l'(r)$ by solving the radial equation, Eq.~(\ref{eq:radial}), with {\tt NDSolve} using the {\tt StiffnessSwitching} option;
  \item in the far-field, extract the coefficients $\Aout$ and $\Ain$ by matching the numerical solution onto the generalized series solutions 
 \beq
 u_l^+(r) = \exp(i \omega r_\ast) \sum_{j=0}^{n_1} b_j r^{-j}, \quad \quad u_l^{\mp}(r) = u_l^{\pm\ast}(r) ,
 \eeq
 by inverting the equations
 \beq
 \begin{pmatrix}  
 u_l^+(\rmax) & u_l^-(\rmax) \\ {u_l^+}'(\rmax) & {u_l^-}'(\rmax) 
 \end{pmatrix}
 \begin{pmatrix} \Aout \\ \Ain \end{pmatrix} =
  \begin{pmatrix} u_l(\rmax) \\ u'_l(\rmax) \end{pmatrix} \, ;
 \eeq
 \item apply Eq.~(\ref{eq:S}) to obtain $S_l$.  
\end{enumerate}
Typically, we used the following choice of internal parameters: $\rmin = 0.1 M$, $\rmax = 1500M$, $n_0 = 20$ and $n_1 = 15$. 

 \subsubsection{Geodesic phase shifts\label{subsec:geophase}}
 
As a consistency check on our numerical scheme, we also computed approximate phase shifts from the geodesic deflection function (see Sec.~\ref{subsec:geodesics}). First, we calculated $\Thetageo(b)$ numerically on a linearly-spaced grid across the domain $0 \le b \le \bmax$ and fitted the data with an interpolating function. Then, the phase shifts $\deltageo$ were obtained from 
\beq
\deltageo = \frac{1}{2} \omega \int_0^b \Thetageo(b') db'  + \chi_0 ,  \label{eq:deltageo}
\eeq
where $b = (l+1/2) / \omega$.
Here $\chi_0$ is an integration constant, which does not affect the cross section $|\hat{f}(\theta)|^2$. It can be fixed by matching to the numerical results in the weak-field regime ($b \gg M$) if necessary.

In the Schwarzschild spacetime, the deflection angle is known in closed form in terms of elliptic integrals. In the weak-field, one may use a power-series expansion (see e.g. Eq.~(33) in Ref.~\cite{Batic:2014loa}),
 \beq
 -\Thetageo(b) = \frac{4M}{b} + \frac{15\pi M^2}{4b^2} + \frac{128M^3}{3b^3} + \frac{3465\pi M^4}{64 b^4} + \frac{3584 M^5}{5b^5} + O(b^{-6}).
 \eeq
 Upon integrating, one obtains
 \beq
  \deltageo = M \omega \left( - 2 \ln (b/M) + \frac{15 \pi M}{8 b} + \frac{32M^2}{3b^2} +\frac{1155\pi M^3}{128b^3} + \frac{448 M^4}{5 b^4} + O(b^{-5}) \right) + \chi_1, \label{eq:deltageo-weak}
 \eeq
 where $\chi_1$ is the constant of integration. 

 \subsubsection{Series convergence}
 For Coulomb scattering, it is well-known that scattering coefficients $S_l$ do not approach unity as $l \rightarrow \infty$, due to the long-range nature of the field ($\sim 1/r$); thus the Coulomb version of Eq.~(\ref{eq:fscat}) is not strictly convergent. Since gravity is also long-ranged ($\sim 1/r$), a similar lack of convergence is expected, and indeed, it is manifest in the logarithmic term in the phase in Eq.~(\ref{eq:deltageo-weak}). The lack of convergence is related to the physical divergence of the cross section in the $\theta \rightarrow 0$ limit. Fortunately, there is a practical remedy, introduced in Ref.~\cite{Yennie:1954zz} in the 1950s. We use
 \beq
\fscat(\theta) = (1 - \cos \theta)^{-n} \fscat^{(n)}(\theta) , \quad \quad  n \in \mathbb{N} ,
\eeq
where $\fscat^{(n)}$ is a `reduced' series of the form 
\beq
\fscat^{(n)}(\theta) = \sum_{l = 0}^\infty (2 l + 1) c_l^{(n)} P_l(\cos \theta) ,
\eeq
whose series coefficients $c_l^{(n)}$ satisfy the recurrence relation 
\beq
(2 l + 1) c_l^{(n+1)} = (2 l + 1) c_l^{(n)} - (l+1) c_{l+1}^{(n)} - l c_{l-1}^{(n)}   ,
\eeq
with $c_l^{(0)} = S_l - 1$. We find that the $n=2$ reduced series is sufficiently convergent for numerical evaluation. 

 \subsubsection{Scattering cross sections\label{subsec:geoscat}}
In Sec.~\ref{sec:results} we present scattering cross sections $d\sigma/d\Omega=|\fscat(\theta)|^2$ labelled either `partial-wave' or `geodesic'. The former (`partial-wave') were obtained by summing the reduced partial-wave series constructed from the first 300 scattering coefficients $S_l$ found by solving the wave equation (see Sec.~\ref{subsec:numerical}). To improve accuracy in the small-angle regime ($\theta \lesssim 15^\circ$), we extended the series beyond $l=300$ using the weak-field phase shifts (\ref{eq:deltageo-weak}) in the regime $300 < l \lesssim 600$.  The latter (`geodesic') were found by summing the reduced series of `geodesic' phase shifts, determined numerically from Eq.~(\ref{eq:deltageo}), without solving the wave equation directly. The `geodesic' results are included for two reasons: as a consistency check on our principle results, and to demonstrate the utility of the geodesic phase shifts for short-wavelength scattering ($M\omega \gg 1$).

\section{Results\label{sec:results}}
Here we present a selection of numerical results illustrating some general features of rainbow scattering by spherically-symmetric compact objects.

\subsection{Scattering coefficients for compact bodies and black holes}
\begin{figure}
 \includegraphics[width=0.75\textwidth]{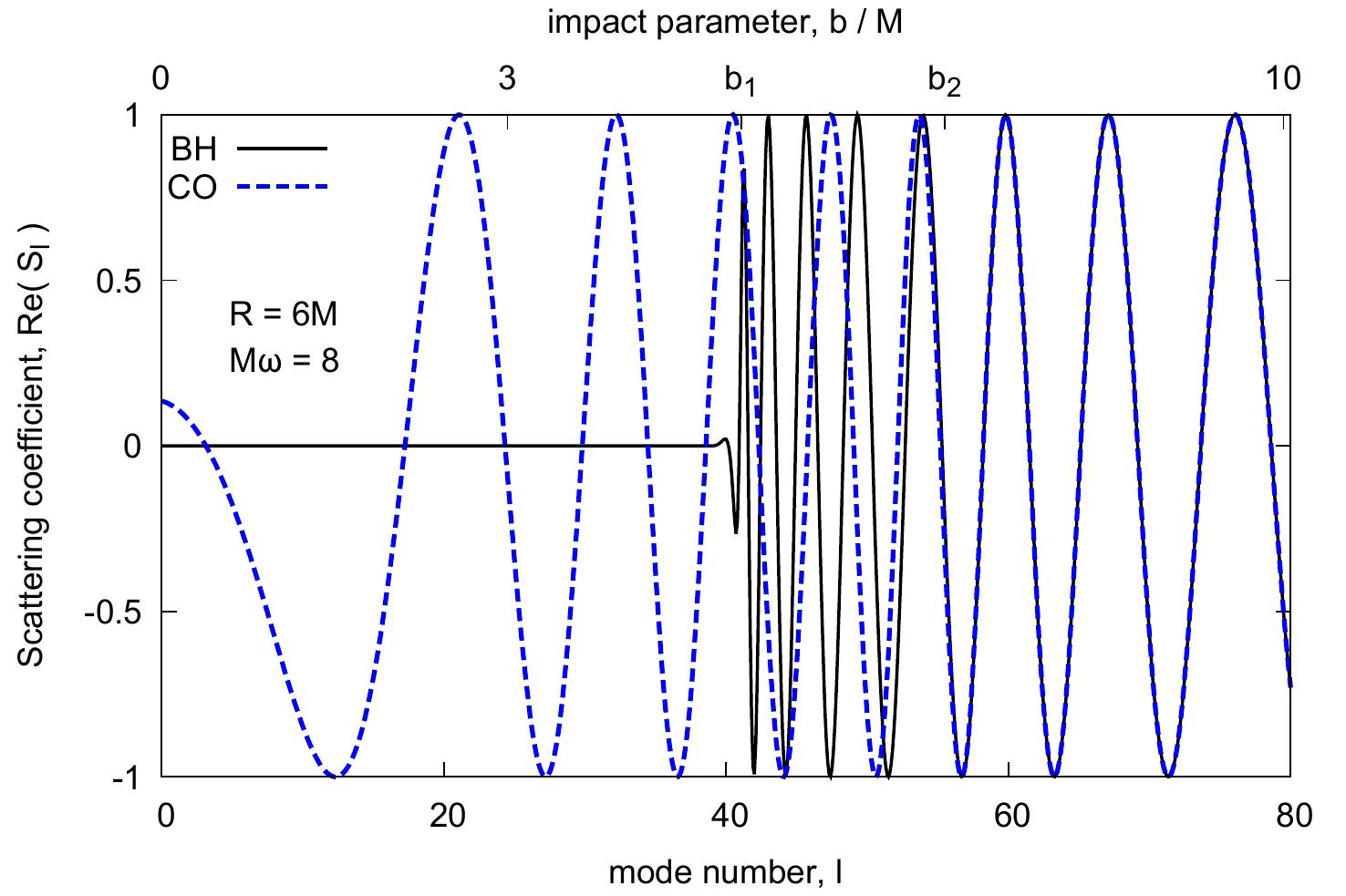}
\caption{
Scattering coefficients $S_l$ for a black hole [solid] and a compact body [blue, dashed] of the same mass. Here $b_1 = \sqrt{27}M$ is the impact parameter associated with the black hole photon orbit, and $b_2$ is the impact parameter associated with the ray that grazes the surface of the compact body. The two x-axes are related by the linear relation $b / M= (l+1/2) / M\omega$.
}
\label{fig:scatco}
\end{figure}

Figure \ref{fig:scatco} shows typical scattering coefficients $S_l$ [see Eq.~(\ref{eq:S})] in the black hole and compact body cases. The similarities and differences can be understood via the semi-classical picture, in which a partial wave with mode number $l$ is associated with a ray with impact parameter $b = (l+1/2) / \omega$. There are two key values of the impact parameter: $b_1 = \sqrt{27}M$ for the ray that asymptotes to the photon orbit at $r=3M$ (where it exists), and $b_2(R,M)$ for the ray that grazes the surface of the compact body. For $b > b_2$ the black hole and compact body scatter in a similar way ($S_l^{CO} \approx S_l^{BH}$), as the associated rays remain in the vacuum exterior. For $b < b_2$, rays pass through the interior of the compact body; and for $b < b_1 = \sqrt{27M}$, rays are absorbed by the black hole and hence $S_l^{BH} \rightarrow 0$, as shown in Fig.~\ref{fig:scatco}.

\subsection{Rainbow scattering: $R > 3M$}
\begin{figure}
\subfloat[Rainbow scattering for $M\omega=8$ and $R=6M$]{\label{subfig:rainbowR6}%
  \includegraphics[width=0.75\textwidth]{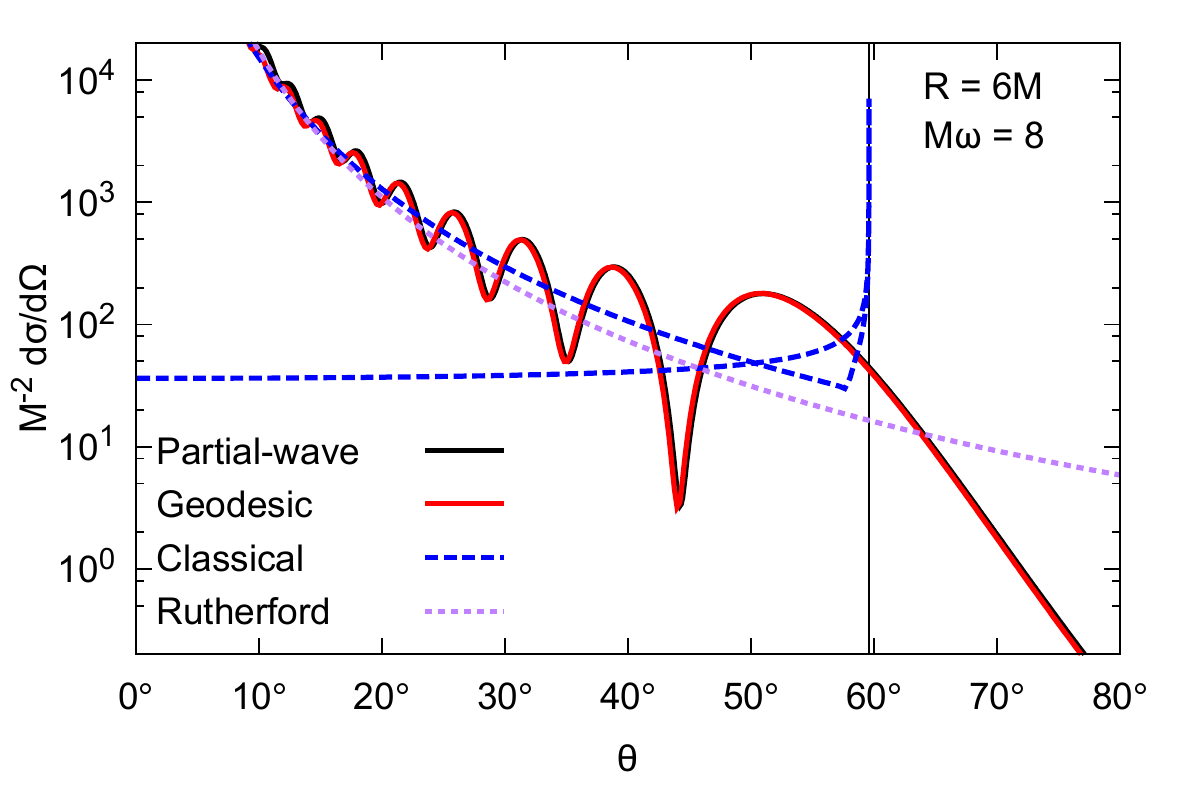}
} \\
\subfloat[Rainbow scattering for $M\omega=8$ and $R=10M$]{\label{subfig:rainbowR10}%
  \includegraphics[width=0.75\textwidth]{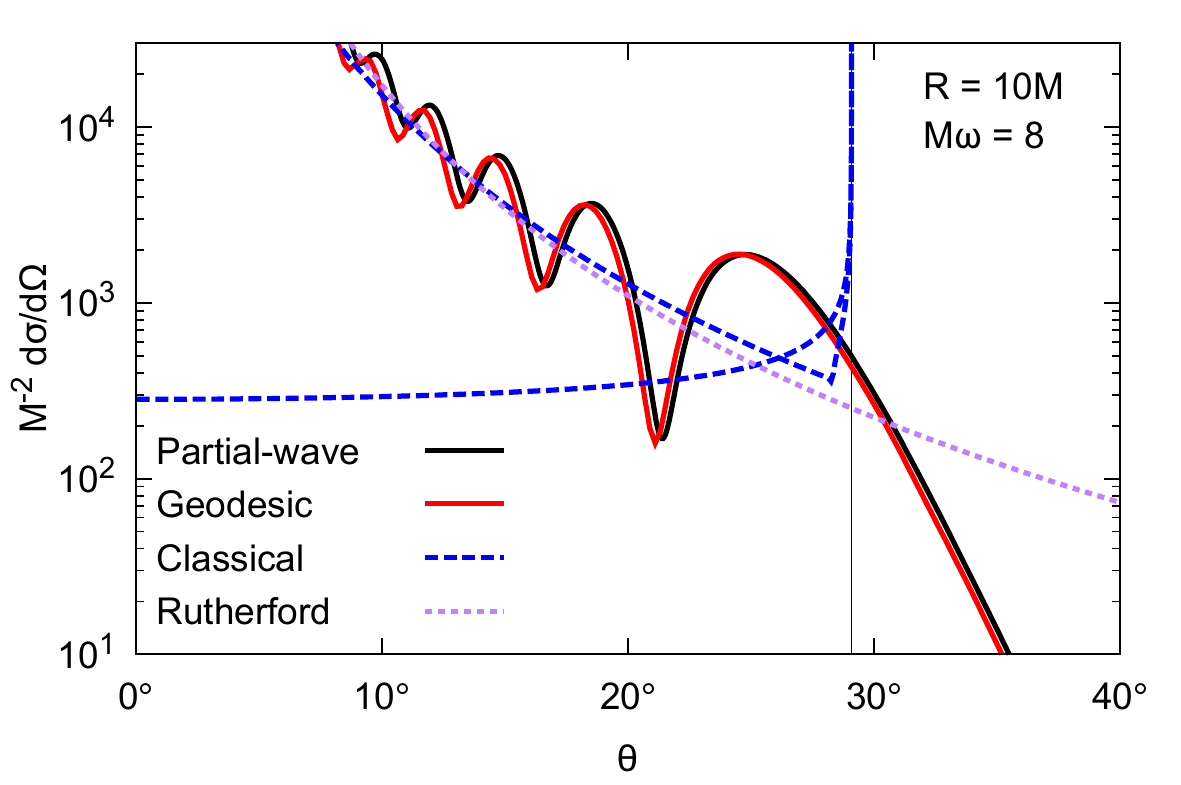}
}
\caption{
Rainbow scattering for compact bodies of radius $R = 6M$ and $R = 10M$, for monochromatic waves of angular frequency $\omega = 8 M^{-1}$. The solid lines show the partial-wave cross section computed from wave-equation phase shifts (black) and geodesic phase shifts (red). The dashed line (blue) shows the classical cross section, Eq.~(\ref{eq:classical}), calculated from the geodesic deflection function $\Thetageo(b)$ of Fig.~\ref{fig:deflection}. The dotted line (purple) shows the Rutherford cross section, $\sin^{-4}(\theta / 2)$, for comparison. A vertical line indicates the geodesic rainbow angle at (a) $\thetabow = 59.6^\circ$ and (b) $\thetabow = 29.1^\circ$ where $\Thetageo' = 0$. 
}
\label{fig:rainbow}
\end{figure}

Figure \ref{fig:rainbow} shows typical rainbow scattering cross sections at the wave frequency $\omega = 8 M^{-1}$, for uniform density stars of radii $R = 6M$ and $R = 10M$. We observe a standard Airy-type oscillation in $d\sigma/d\Omega$ to the left of a rainbow angle $\thetabow$, followed by exponential suppression in the shadow region to the right of $\thetabow$. As expected, the `primary' rainbow peak appears at an angle somewhat smaller than $\thetabow$. The angular width of the oscillations decreases as $\omega$ increases. Increasing the wave frequency moves the primary peak further towards the rainbow angle. This chromatic effect is analogous to that responsible for coloured bands in the optical rainbow. There is good agreement between the `partial-wave' and `geodesic' cross sections (see Sec.~\ref{subsec:geoscat}).

 Heuristically, the cross section can be viewed as a regular interference effect superimposed on a classical cross section. The two branches of the classical cross section (blue in Fig.~\ref{fig:rainbow}) correspond to deflection angles from either side of the minimum in $\Thetageo$. The branch that is regular as $\theta \rightarrow 0$ comes from low-$l$ waves passing into the compact body; the branch that is divergent as $\theta \rightarrow 0$ comes from large-$l$ waves experiencing weak-field scattering $\Thetageo(b) \sim - 4M / b$. The plot shows that the magnitude of the regular branch approximately determines the amplitude of the interference oscillations around the irregular branch.

\subsection{The Airy approximation for rainbows\label{subsec:airy}}
In rainbow scattering, the spectral components of an incident wave are separated by wavelength. That is, the scattering angle of the primary peak is a function of wavelength. For longer wavelengths, $M\omega \sim 0.1$ -- $10$ we can compute cross sections numerically. For short wavelengths, $M\omega \gtrsim 10$ it is more convenient to use semiclassical approximations, such as Eq.~(\ref{eq:airy}). 

\begin{figure}
\subfloat[$R=6M$ and $R=10M$ with $M\omega=8$.]{\label{subfig:airy1}%
  \includegraphics[width=0.52\textwidth]{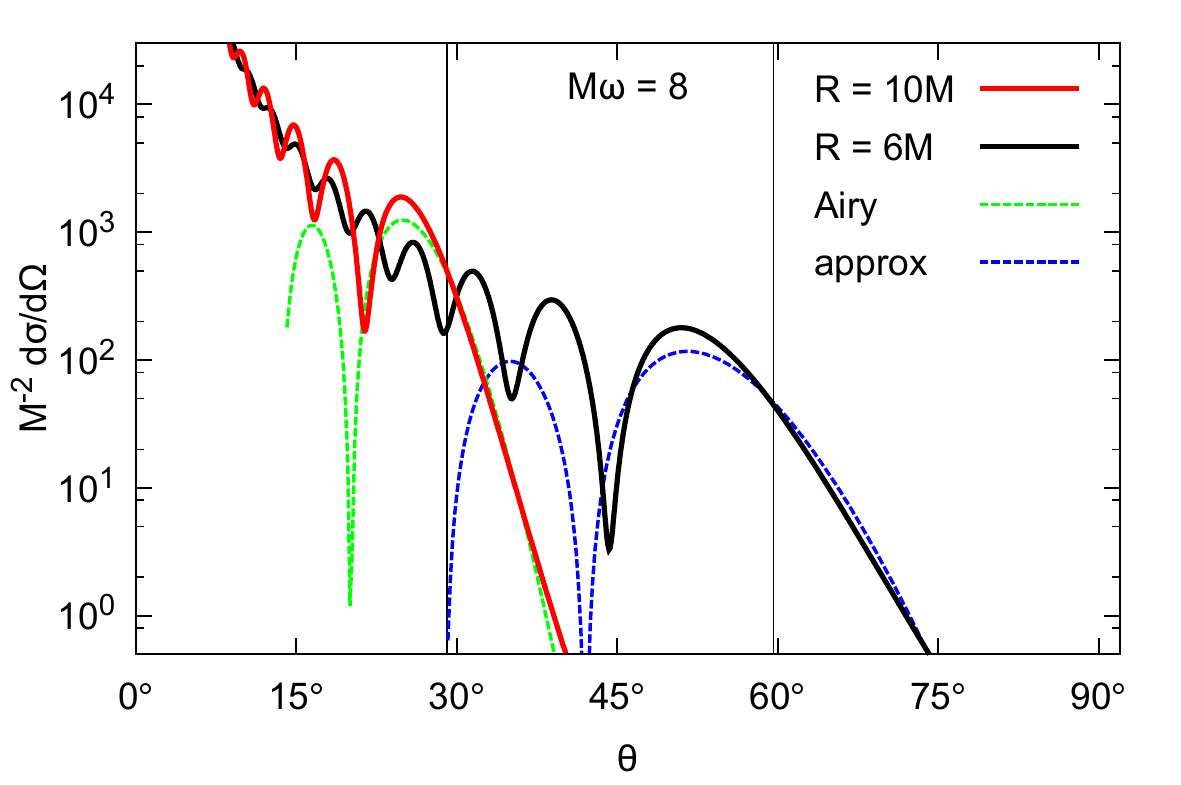}%
}
\subfloat[$R=10M$ with $M\omega = 10, 30, 50$ and $100$.]{\label{subfig:airy2}%
  \includegraphics[width=0.52\textwidth]{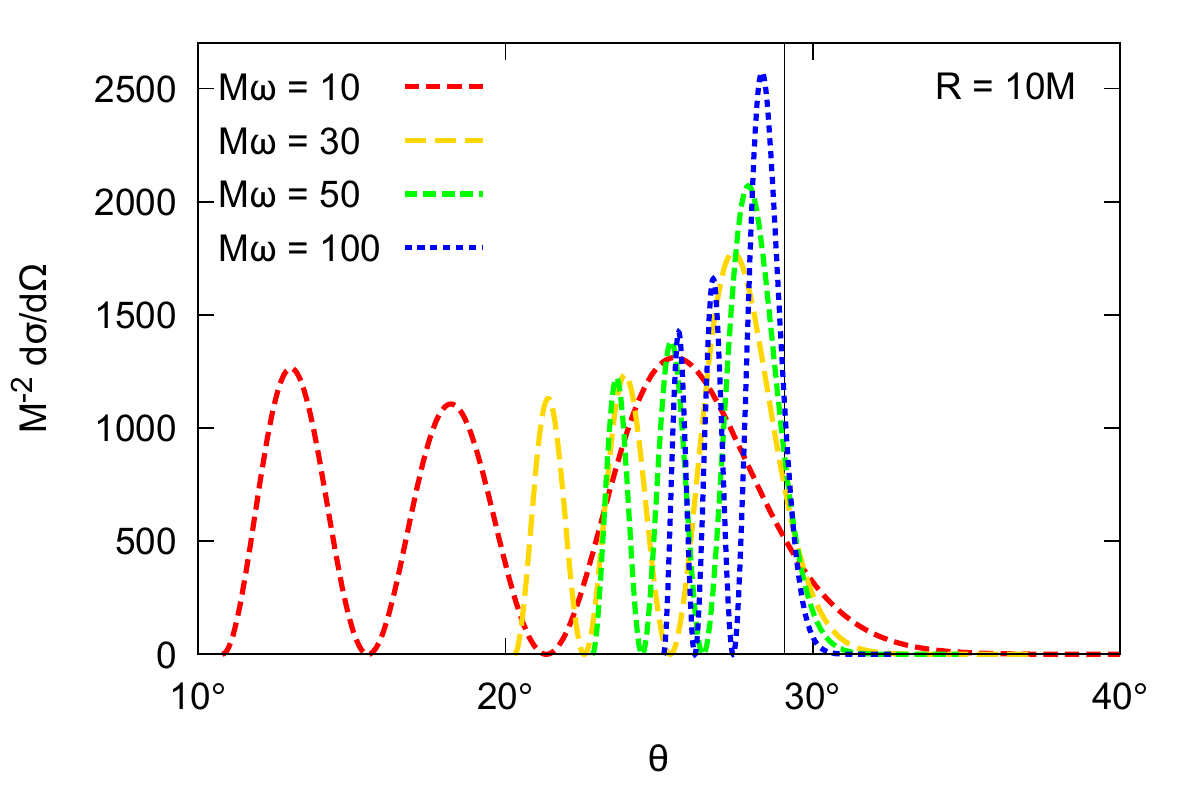}%
}
\caption{
The Airy approximation for rainbow scattering. The dashed lines show the Airy approximation, Eq.~(\ref{eq:airy}), using the geodesic rainbow parameters $\bbow$, $\thetabow$ and $\Theta''_r$.  (a) The solid lines show the numerically-determined cross sections for compact bodies of radius $R = 6M$ (black) and $R = 10M$ (red), and the vertical lines show the rainbow angle $\theta_r$ for each case. (b) For shorter wavelengths, the primary peak lies closer to $\Theta_r$, creating a rainbow effect.
}
\label{fig:airy}
\end{figure}

Figure \ref{subfig:airy1} compares the Airy approximation, Eq.~(\ref{eq:airy}) computed with the geodesic parameters $\bbow$, $\thetabow$ and $\Theta_r''$, with numerical data. We see that, at $M\omega = 8$, the Airy approximation captures, to a reasonable accuracy, the angle, width and intensity of the primary peak. As expected, it is most accurate for $\theta \sim \theta_r$. 
As the Airy approximation (\ref{eq:airy}) was derived with semiclassical methods, we should expect it to become increasingly accurate at short wavelengths, $M \omega \rightarrow \infty$. Figure \ref{subfig:airy2} shows the primary rainbow peak for various couplings $M \omega$ in the case $R = 10M$. The primary peak appears at angle $\theta_{\text{peak}} \approx \thetabow - 0.237 [\lambda^2 \Theta_r'']^{1/3}$ (where $\lambda = 2\pi / \omega$). Thus, the peak approaches the rainbow angle as $\lambda \rightarrow 0$. The intensity increases in proportion to $(M\omega)^{1/3}$.

\subsection{Wide-angle rainbows and enhanced glories: $R \sim 3.5M$}

The rainbow angle $\thetabow$ increases as the compactness parameter $R/M$ decreases. For $R = 3.5M$, the geodesic rainbow angle is $\thetabow \approx 189.4^\circ$. Naively, one would expect two consequences. Firstly, as there is no `shadow zone' in this case, one would expect significant scattering through all angles. Secondly, one would expect an enhanced glory effect, due to a coalescence of the two types of divergence ($\sin \theta = 0$ and $\Theta' = 0$) in the classical cross section, Eq.~(\ref{eq:classical}). Our results show that these expectations are well-founded. 

\begin{figure}
\includegraphics[width=15.0cm]{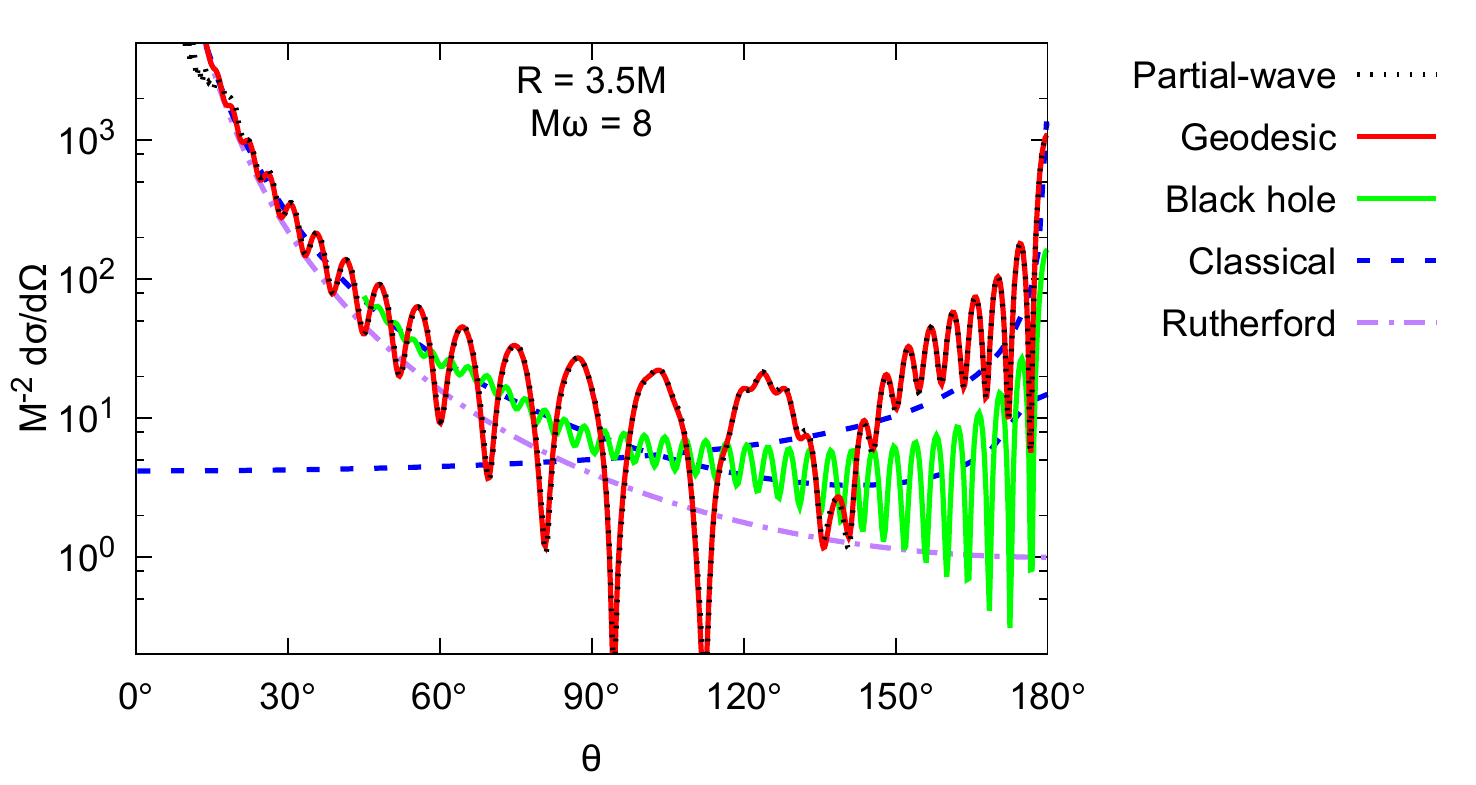}
\caption{
Enhanced glory scattering for a very compact body $R = 3.5M$ caused by the large rainbow angle, $\thetabow \approx 189.4^\circ$. The `wide' and `narrow' oscillations arise from the $b < \bbow$ and $b > \bbow$ branches of the deflection function. The compact body (red) scatters more flux through wide angles than a black hole (green) of the same mass. (See also caption of Fig.~\ref{fig:rainbow}).
}
\label{fig:rainbow-glory}
\end{figure}

Figure \ref{fig:rainbow-glory} shows the scattering cross section for a very compact body with $R = 3.5M$. We note that the `geodesic' cross section is a very good approximation to the `partial-wave' cross section in this case (Sec.~\ref{subsec:geoscat}). The cross section exhibits wide and narrow orbiting oscillations. These may be understood by examining the two branches of the classical cross section, from $b < \bbow$ (interior) and $b > \bbow$ (exterior). At small angles, the exterior branch is dominant, and the interior branch -- associated with rays that pass through the body -- creates interference oscillations. For $\theta \gtrsim 105^\circ$, the interior branch becomes dominant, and the interference oscillations arise from the exterior branch.

In comparison with the black hole case (Fig.~\ref{fig:rainbow-glory}, green) the compact body cross section shows much more pronounced orbiting oscillations, a significantly higher flux at large angles, and a more intense glory. For $M\omega=8$, the intensity of the glory peak at $\theta = 180^\circ$ exceeds the cross section at all angles beyond $\sim 21^\circ$. Semiclassical theory implies that the glory intensity will increase linearly with $\omega = 2\pi / \lambda$ \cite{Matzner:1985rjn}.

\subsection{Ultra-compact bodies with light rings: $R<3M$}

\begin{figure}
\includegraphics[width=12.0cm]{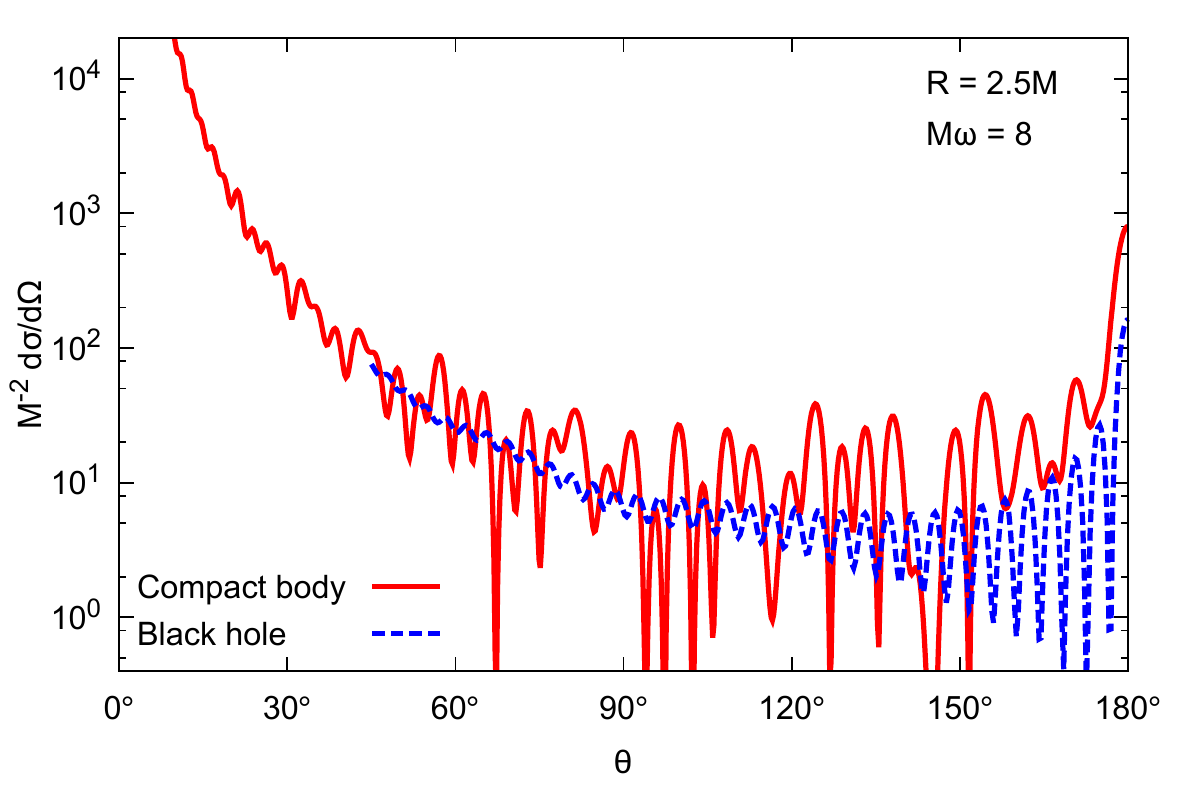}
\caption{
Orbiting and interference for ultra-compact body with a light-ring, $R = 2.5M$ (red), compared with the black hole case (blue, dashed). 
}
\label{fig:rainbow-lightring}
\end{figure}

Figure \ref{fig:rainbow-lightring} shows an example of scattering from an ultra-compact body whose radius $R=2.5M$ is smaller than the light-ring radius $r_c = 3M$. In this case, the deflection function (Fig.~\ref{fig:deflection})  diverges at $b = b_c = \sqrt{27} M$, as in the black hole case. Unlike the black hole case, the deflection function is well defined for $b < b_c$, as waves can pass into the body and come out again.

The compact-body cross section (Fig.~\ref{fig:rainbow-lightring}, red) exhibits orbiting oscillations which are much less regular in appearance than in the black hole case. This is because the scattering amplitude $\fscat(\theta)$ is the sum of contributions from the interior ($b < b_c$) and exterior ($b > b_c$) branches of the deflection function (Fig.~\ref{subfig:deflection}) at the angles $\theta$, $2 \pi - \theta$, $2\pi + \theta$, etc. Waves passing inside the light-ring radius can be scattered through arbitrarily large angles. 

For $R < 3M$, the `geodesic' phase shifts (Sec.~\ref{subsec:geophase}) are no longer good approximations for the partial-wave phase shifts in the regime $b \sim b_c$. The WKB approximation breaks down when multiple turning points are close together, and an additional phase shift is accrued in transitioning from $b > b_c$ to $b < b_c$. It seems that this phase shift cannot be straightforwardly inferred from the geodesic analysis.

\section{Discussion\label{sec:discussion}}
In this work we have investigated the time-independent scattering of planar waves by a spherically-symmetric compact body. We have shown that the key features of scattering are related to the key properties of the geodesic deflection function $\Thetageo(b)$ (see Fig.~\ref{fig:deflection} and Fig.~\ref{fig:potential}). For compactness ratios $R/M > 3$, a caustic forms (Fig.~\ref{subfig:caustic}), leading to rainbow scattering, i.e., regular interference oscillations in the cross sections (Figs.~\ref{fig:rainbow}) which have more than a passing resemblance to those in nuclear rainbow scattering \cite{Goldberg:1974zza, Khoa:2006id}. In Fig.~\ref{fig:airy} we showed that the rainbow oscillations are well-modelled by Airy's approximation (\ref{eq:airy}) using the geodesic parameters $\theta_r$, $b_r$ and $\Theta_r''$, in the semiclassical regime $M\omega \gg 1$ . The rainbow angle $\thetabow$ increases as the body becomes more compact and, for $R/M = 3.5$, $\thetabow$ exceeds $180^\circ$. In Fig.~\ref{fig:rainbow-glory} we showed that this leads to an enhanced glory in the backward direction. Finally, in Fig.~\ref{fig:rainbow-lightring}, we showed that ultra-compact bodies with $R/M < 3$ generate complex scattering patterns, due to the interference between rays that pass close to the light-ring and those that pass into the body itself.

We have shown here that the scattering pattern from a compact body is rather different to that from a black hole. In the former case, the stationary point in $\Thetageo(b)$ generates a rainbow; in the latter case, a divergence in $\Theta(b)$ associated with the light-ring generates spiral scattering and a glory. In considering more exotic compact bodies (boson stars, wormholes, etc.), one should give thought to which of these effects will occur. It would be particularly interesting to investigate scenarios with multiple stationary points of $\Theta(b)$, or scenarios in which both effects occur. This may be possible in (e.g.) the hairy black hole scenarios recently described in Refs.~\cite{Herdeiro:2014goa,Herdeiro:2015waa,Herdeiro:2016tmi}.

Let us address some of the limitations of the model. 
In this work we have considered only a scalar field. What differences are anticipated for a gravitational wave? Firstly, it is known that for massless fields of non-zero spin (such as the neutrino, electromagnetic or gravitational fields) the backward glory will resemble a ring rather than a bright spot, as parallel-transport in a spherically-symmetric spacetime leads to perfect destructive interference at $\theta = 180^\circ$, i.e., $\fscat(\pi) = 0$ \cite{Matzner:1985rjn}. (A subtlety in the gravitational-wave case is that the central mass also generates a helicity-reversing scattering amplitude $\hat{g}(\theta)$ \cite{Handler:1980un,Futterman:1988ni, Dolan:2007ut, Dolan:2008kf} and $\hat{g}(\theta) \neq 0$ in the backward direction.) It is plausible that spin-related interference effects will occur at other angles within the rainbow; this deserves further investigation. Secondly, gravitational-wave detectors measure amplitude, rather than intensity. Thus it is the scattering amplitude $\fscat(\theta)$ rather than the cross section $d\sigma/d\Omega = |\fscat(\theta)|^2$ that should be the central object of interest in any study of gravitational waves.

We have considered the simplest model available: the spacetime of a uniform density star \cite{Schwarzschild:1916}. With a different spherically-symmetric density profile we would expect the rainbow effect to survive, but the key parameters $\bbow$, $\thetabow$ and $\Theta''_r$ to be shifted. A more realistic model of a body with mass and spin multipole moments would modify the effect more fundamentally, for example, by reducing the regularity of the Airy oscillations. For example, a rapidly-spinning body would generate frame-dragging, distinguishing between prograde and retrograde rays. Again, this is a subject for future investigation. 

An open question is whether short-wavelength effects such as rainbows are at all relevant for the nascent science of gravitational-wave astronomy. Let us examine three aspects of this question. 

(1) Can the gravitational-wave wavelength ever be sufficiently short in comparison to the dimensions of a compact scatterer? Let us suppose the gravitational wave was generated by the $l=2$ quasinormal mode frequency of a Schwarzschild black hole, with $\omega \approx 0.374 M_{bh}^{-1}$, and that it impinged upon a neutron star of mass $M = 1.4M_{\odot}$. In this scenario, $M \omega \sim 1.52 (M_{bh}/M_{\odot})^{-1}$ and $R/M \sim 6$; thus, in this case the validity of the semiclassical assumption $\omega R \gg 1$ is somewhat questionable. However, if one replaces the neutron star with a typical white dwarf, $R / M \sim 9.4\times10^3$, then the semiclassical assumption is well justified. 

(2) Is time-independent scattering relevant in gravitational-wave scenarios? The first observed gravitational events (black hole binary signals GW150914 and GW151226) are short-lived chirps ($<1$s). On the other hand, a key target for (future) space-based detectors (e.g.~LISA) are long-lasting ($\sim 1$ yr), low-frequency quasi-periodic signals from extreme mass ratio inspirals. We take the view that time-independent scattering is worth studying, for two reasons. First, as GW radiation is coherent, so interference effects are relevant in principle, and time-independent scattering offers a `scaffold' for understanding interference effects in time-dependent scattering. Second, comparisons with scattering scenarios in other parts of physics -- for example, nuclear rainbow scattering \cite{Khoa:2006id} -- can lead to deeper physical understanding. 

(3) Are compact-body scattering scenarios sufficiently intense to be observable with current technology? This seems unlikely. In this work we have considered the secondary scattering of some pre-generated wave by an isolated body; whereas the experimental focus is rightly on the direct observation of GWs from the loudest, cataclysmic events such as binary mergers and supernovae. We take the view that, even if secondary scattering effects are not detectable in the near-future, it is still interesting to know that such effects exist in principle, and that such effects clearly discriminate between the black hole and compact body scenarios. It is not beyond conception that an advanced civilization could use gravitational-wave rainbow scattering measurements to probe the internal structure of the neutron star in much the same way as nuclear rainbow scattering measurements have been used, since the 1970s, to examine the structure of the nucleus \cite{Khoa:2006id}. 

The lensing of gravitational waves by foreground matter distributions has received recent attention. Lensing affects the apparent luminosity of gravitational waves \cite{Dai:2016igl, Bertacca:2017vod}, and induces timing delays \cite{Baker:2016reh}. It is plausible that interference effects, such as rainbows, may also be relevant. For example, the GW signal from an intermediate or extreme mass-ratio inspiral will slowly sweep across a waveband, increasing in frequency over time as the orbit tightens. As the frequency increases, an observer at fixed angle can move from a n interference peak to a trough (see Fig.~\ref{subfig:airy2}). In addition, the profile of the signal in the time-domain will change on passage through a caustic \cite{Dolan:2011fh, Zenginoglu:2012xe, Dai:2017huk}. Further study of small-angle rainbow scattering from mass distributions with large compactness ratios $R/M$ seems desirable.

Is gravitational rainbow scattering from compact objects relevant in the electromagnetic sector? Here we should be cautious, because there are a host of electromagnetic processes in astrophysics which could inhibit or obscure the effect. We note from Fig.~\ref{fig:deflection} that the rainbow angle is associated with a null geodesic that passes through the outer part of the compact body. Thus, at the very least, one should consider absorption and/or re-scattering of rays near the surface the compact body. However, if the compact body were instead some dark-matter distribution, then attenuation may be negligible. One could also consider the scattering of neutrinos, or some other weakly-interacting field \cite{Jaeckel:2010ni,Arias:2012az}. If the fields can penetrate the outer layers of the object, and if the field is sufficiently coherent, then we should expect a rainbow effect along the lines described here.

Finally, to what extent could the spacetime geometry of our nearest massive body act to focus gravitational radiation? In the geometric-optics regime, the Sun works as a gravitational lens, generating an `Einstein ring' of angular radius $\theta_{\text{ring}} \sim \sqrt{4M / d}$, where $d$ is the distance from the observer to the solar system centre (e.g.~$\theta_{\text{ring}} \sim 41$ arcseconds for observers near Earth). Relatedly, a caustic associated with the Sun's deflection function will lie at a distance of $d \sim \bbow^2 / 4M$ from the solar system centre (see Fig.~\ref{subfig:caustic}). A uniform density model with $\bbow \sim R_{\odot} \approx 6.96 \times 10^8\,\text{m}$ leads to a very crude estimate of $d \sim 550 \, \text{au}$ (astronomical units); whereas for a centrally-dense radial profile, $d$ will be somewhat smaller. However, in Ref.~\cite{Bontz-1981} it was shown that the large intensity increases associated with a caustic are reduced by diffraction effects if the wave frequency is less that $\omega_c \sim (10^{-1} \pi M)^{-1}$. For the Sun, the critical frequency of $\omega_c \sim 10^4 \text{s}^{-1}$ is higher than the typical frequency of gravitational waves from astrophysical sources; thus placing a GW detector near the caustic is unlikely to bring scientific benefits.

\acknowledgments
S.D.~acknowledges financial support from the Engineering and Physical Sciences Research Council (EPSRC) under Grant No.~EP/M025802/1 and from the Science and Technology Facilities Council (STFC) under Grant No.~ST/L000520/1. T.S. acknowledges financial support from EPSRC through the University of Sheffield Doctoral Training Partnership Scholarship.
\bibliographystyle{apsrev4-1}
\bibliography{refs}

\end{document}